\documentclass[aps,prb,twocolumn,superscriptaddress]{revtex4-1}
\usepackage{graphicx}
\usepackage{epstopdf}
\usepackage{amsmath}
\usepackage{amsfonts}
\usepackage[usenames,dvipsnames]{color}
\usepackage{bm}
\usepackage{amsmath}
\usepackage{changes}
\usepackage[colorlinks=true,linkcolor=blue,urlcolor=blue,citecolor=blue]{hyperref}
\usepackage{lineno, blindtext}

\definecolor{pink}{rgb}{1,1,0} % color values Red, Green, Blue
\definecolor{red}{rgb}{1,0,0}
\definecolor{yellow}{rgb}{1,1,0}
\definecolor{orange}{rgb}{1,0.5,0}
\definecolor{green}{rgb}{0,1,0}
\definecolor{blue}{rgb}{0,0,1}
\definecolor{white}{rgb}{1,1,1}
\definecolor{purple}{rgb}{0.5,0,0.5}

\begin{document}
\title{Intrinsic topological magnons in arrays of magnetic dipoles}
\author{{ Paula Mellado$^{1,2}$}\\
{\small \em $^1 $School of Engineering and Sciences, 
	Universidad Adolfo Ib{\'a}{\~n}ez,
	Santiago, Chile\\
	\small \em $^2 $CIIBEC, Santiago, Chile.}}

\begin{abstract}
We study a simple magnetic system composed of periodically modulated magnetic dipoles with an easy axis. Upon adjusting the modulation amplitude alone, chains and two-dimensional stacked chains exhibit a rich magnon spectrum where frequency gaps and magnon speeds are easily manipulable. The blend of anisotropy due to dipolar interactions between magnets and geometrical modulation induces a magnetic phase with fractional Zak number in infinite chains and end states in open one-dimensional systems. In two dimensions it gives rise to topological modes at the edges of stripes. Tuning the amplitude in two-dimensional lattices causes a band touching, which triggers the exchange of the Chern numbers of the volume bands and switches the sign of the thermal conductivity.  

\end{abstract}

\maketitle

%\begin{linenumbers}
\section{Introduction}
\label{sec:intro}
The latest experimental findings associated with twisted heterostructures \cite{cao2018unconventional}, demonstrated that handling a single geometrical parameter in a system is a simple yet highly effective strategy for manipulating electric correlations and realizing diverse electronic, magnetic, and topological phases. In the realm of magnetic natural and artificial matter, the recent proposal on Moire magnets \cite{hejazi2020noncollinear}, magnetic counterpart to twistronic, promises to complement and enlarge the current possibilities for manipulation of the flow of spin angular momentum with and without an accompanying charge current \cite{hirsch1999spin, tong2018skyrmions,li2020moire}. 
Currents of spins can be produced by the transport of electrons or, without their aid, by the collective propagation of coupled precessing spins or spin waves \cite{van1958spin}. In magnetically ordered systems, based on magnetic insulators, ferromagnetic metals or heterostructures, spin-wave excitations, and their quantized versions, magnons \cite{cornelissen2015long,chumak2015magnon} can be controlled by varying the layer thicknesses and applying external fields \cite{gartside2020current,wang2018topological,krawczyk2014review,fan2020manipulation}. In artificial magnonic crystals \cite{pirmoradian2018topological}, the periodic modulation of the magnetic properties can be engineered, which makes possible the formation of magnonic bands separated by gaps in one-dimensional, two-dimensional, and bicomponent structures \cite{iacocca2016reconfigurable}. The band's gaps in these structures can be manipulated by combining changes in the crystal parameters, modification of crystal materials, and the tuning of direction and amplitude of magnetic fields \cite{krawczyk2014review,zang2011dynamics,diaz2019topological,diaz2020chiral,kim2019tunable}. 

In addition to gap manipulation and the reduction of heat associated with dissipation, the unidirectional transmission of information using spin waves is another aspect being considered. In topological magnonics \cite{wang2018topological,zhang2013topological,mook2014edge,chisnell2015topological}, the bulk-edge correspondence \cite{girvin2019modern} ensures that spin waves at the edges of a sample realize chiral propagation along the edges regardless of the specific device geometry. Nontrivial topology is usually accompanied by the emergence of exotic phenomena \cite{kato2004observation,nagaosa2010anomalous,katsura2010theory}. This is the case in $\rm{Lu_2V_2O_7}$, an insulating collinear ferromagnet pyrochlore where spin excitations give rise to the anomalous thermal Hall effect \cite{nagaosa2010anomalous}. In $\rm{Lu_2V_2O_7}$ the propagation of the spin waves is influenced by the antisymmetric Dzyaloshinskii-Moriya (DM) spin-orbit interaction, which plays the role of the vector potential. This is also the case in the compound $\rm{YMn_6Sn_6}$, a metallic system consisting of ferromagnetic kagome planes \cite{li2021dirac}, where the subsequent magnon band gap opening at the symmetry-protected K points is ascribed to DM interactions. Antiferromagnetic materials with honeycomb lattices add to the list of unique crystal structures which realize topological phases \cite{lee2018magnonic,owerre2017topological,PhysRevB.104.L060401}.  

In two dimensions (2D), magnonic crystals can host topological magnons when the system under consideration breaks time and space inversion symmetries \cite{shindou2013topological}. Current evidence shows that this is possible when the system exposes spin-orbit physics on superlattices. In these structures, the common elements usually consist of units cells with the triangular motif and interactions that include all or some of the following elements: DM, anisotropy, exchange interactions, dipolar interactions, and external magnetic fields \cite{mook2014edge,chisnell2015topological,li2021magnonic,nikolic2020quantum}. Control over the band gaps and topological features in these structures include variations in magnetic fields, DM, temperature, and exchange interactions \cite{kim2016realization,pirmoradian2018topological,rau2016spin}. 

In systems where dipolar interactions play a dominant role, several proposals of chiral spin-wave modes in dipolar magnetic films have been put forward \cite{liu2020dipolar,pirmoradian2018topological, shindou2013topological,shindou2013chiral}, when subjected to an external magnetic field. Topological magnons also arise in magnonic crystals with the dipolar coupling truncated after a few neighbor magnets and in the case of antiferromagnetic films in the long-wave limit \cite{liu2020dipolar,pirmoradian2018topological}. In these systems, control over the dipolar magnonic bands is achieved through the application of magnetic fields.  

With all the notable breakthroughs associated to topological magnons, significant progress could be achieved if the shape and the topology of the magnonic bands could be manipulated by precise control of a single geometrical parameter and without the aid of external fields. Here we present a new magnetic system where engineering of the bandgap and Berry curvature is possible by tuning a single intrinsic geometrical parameter. The system consists of two-dimensional lattices in the $x-z$ plane whose building blocks are one-dimensional (1D) lattices of dipoles that extend along the $x$ direction and stack regularly along the $z$ direction. The sites along chains are regularly arranged in a modulated periodic fashion. Dipoles have an easy-axis anisotropy and couple with each other through full anisotropic magnetic dipolar interaction. Equilibrium magnetic states, spin-wave spectrum, and topological features of modulated chains, stripes, and lattices can be explicitly controlled by tuning $\Lambda$, which is defined as the amplitude of the spatial modulation of the chain's sites. The study of the magnon spectrum of chains reveals two topological phase transitions in terms of $\Lambda$. They come along with magnon modes localized at the edges of open systems and jump discontinuities in the Fermi velocity of magnons. The 1D topology is 
inherited by the 2D lattices, which manifest non-trivial topology at $\Lambda>0$ and realize chiral magnonic modes at the edge of stripes. 

The paper is organized as follows. In section \ref{sec:model} we present the model and show the energy spectrum of chains and two-dimensional lattices in terms of $\Lambda$. In section \ref{sec:waves} we show results for the spin-wave spectrum using the Linearized Landau-Lifshitz equations. Section \ref{sec:chern} is dedicated to the study of topological aspects of 1D and 2D systems. We summarize our results in Section \ref{sec:discussion}.
\section{Model}
\label{sec:model}
\begin{figure}
  \includegraphics[width=\columnwidth]{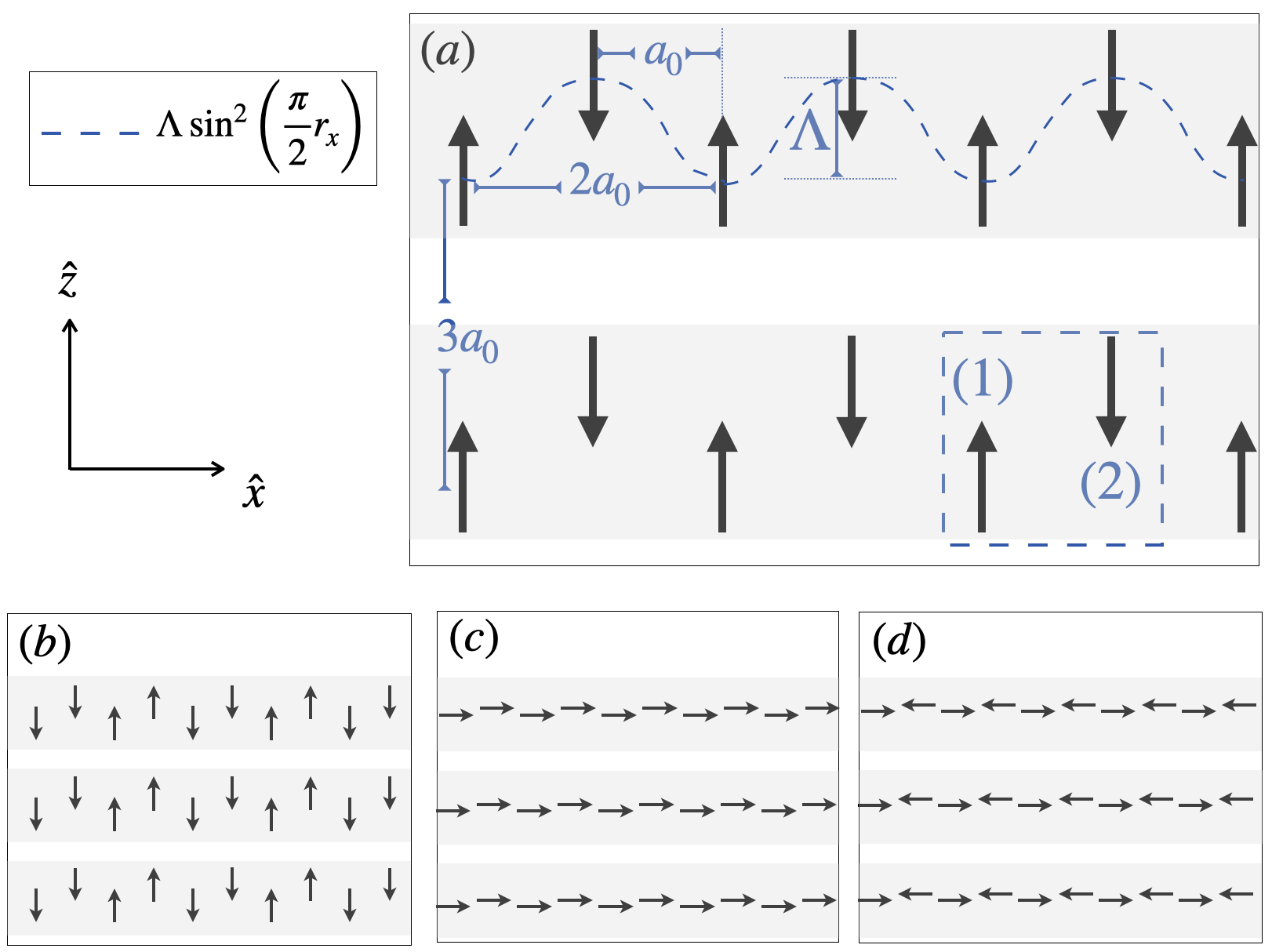}
\caption{Sketches of stripes built out of modulated chains of dipoles. (a) Along each chain, point dipoles are located at  $\Lambda\sin^2(\kappa r_x)$ with $r_x$ the position of dipoles along $\hat{x}$, $\kappa=\pi/2$ the wave vector of the modulation, and $\Lambda$ the amplitude of the modulation as illustrated by the dotted (blue) curve. $a_0$ denotes the distance between two nearest neighbor dipoles along $\hat{x}$. Lattice constant along $\hat{x}$ and $\hat{z}$ are $2a_0$ and $3a_0$, respectively. The dotted square highlights the unit cell, where $(1)$ and $(2)$ denote the two sublattices. (a) and (b) show stripes in the antiferromagnetic and dimer parallel magnetic orders, respectively. (c) and (d) show stripes in the ferromagnetic and antiferromagnetic collinear magnetic orders, respectively. Single chains are highlighted in grey.}
\label{fig:f1}
\end{figure}

The \emph{classical} hamiltonian consists of dipolar interactions and uniaxial anisotropy. In units of Joule $\rm[J]$ it reads 
\begin{eqnarray}
\mathcal{H}&=&\frac{\mathcal{D}}{2} \sum_{i\neq j=1}^n \frac{\hat {\bm  m}_i \cdot\hat {\bm m}_j - 
3 (\hat {\bm m}_i \cdot \hat {\bm{e}}_{ij} )(\hat {\bm m}_j\cdot \hat {\bm{e}}_{ij} )}{|{\bm r}_i -{\bm r}_j |^3}+ \nonumber \\&&  - \,
\frac{{\mathcal K}}{2} \sum_{i=1}^n(\hat {\bm m}_i\cdot \hat {\bm{z}})^2,
\label{eq:Energy}
\end{eqnarray}
where $n$ is the total number of sites and ${\bm r}_i$ denotes the position of a point dipole in a two dimensional array in the $x-z$ plane. $\hat {\bm{e}}_{ij}= ({\bm r}_i -{\bm r}_j ) /|{\bm r}_i -{\bm r}_j |$, and $\mathcal{D}=\frac{\mu_0 m_0^2}{4\pi a_0^3}$ has units of $[\rm Nm]$ and contains the physical parameters involved in the dipolar energy such as $\mu_0$, the magnetic permeability, $a_0$, the distance among nearest neighbour dipoles along the $\hat{x}$ direction, and ${m_0}$, the intensity of the magnetic moments with units $\rm[m^2 A]$. $\mathcal{K}$ is the easy axis anisotropy along the $\hat{z}$ axis and has units of $[\rm Nm]$. Magnetic moments have unit vector
$ \hat{{\bm m}}_i = (\cos\theta_i ,0 ,\sin\theta_i )$. Point dipoles are located at the vertices of modulated chains that extend along the $\hat{\bf x}$ direction as highlighted in Fig.~\ref{fig:f1}. Magnets can rotate in an easy $x-z$ plane described by an azimuthal angle $\theta_i$ which is chosen with respect to the $\hat{z}$ axis. 

The periodic modulation of dipoles in the $x-z$ plane is set by $\Lambda\sin^2(\kappa r_x)$, where $\Lambda$ is the amplitude of the modulation (in units of $a_0$), $r_x$ denotes the position of dipoles along $\hat{x}$ and $\kappa$ is the wave vector of the modulation.  
Modulated chains are one-dimensional lattices with a basis. Depending on $\kappa$, the unit cell contains more than one dipole (each dipole in the unit cell defines a sublattice). Periodic chains are stacked across the $\hat{z}$ axis and are the building blocks of stripes and lattices. Hereafter, we focus on the relevant case of $\kappa=\frac{\pi}{2}$ that corresponds to a lattice with a unit cell containing two dipoles (lattice with a two-point basis). The two sublattices are denoted by (1) and (2), and the unit cell is highlighted by the dotted square in Fig.~\ref{fig:f1}(a). The lattice constants along $\hat{x}$ and $\hat{z}$ axes are $2a_0$ and $3a_0$ respectively, otherwise stated. Systems with $\kappa=\frac{\pi}{4}$ and $\kappa=\frac{\pi}{8}$ are briefly discussed in the supplementary material \cite{supp}. Neither magnon-phonon nor magnon-magnon interactions are considered in this paper.
\begin{figure}
\includegraphics[width=\columnwidth]{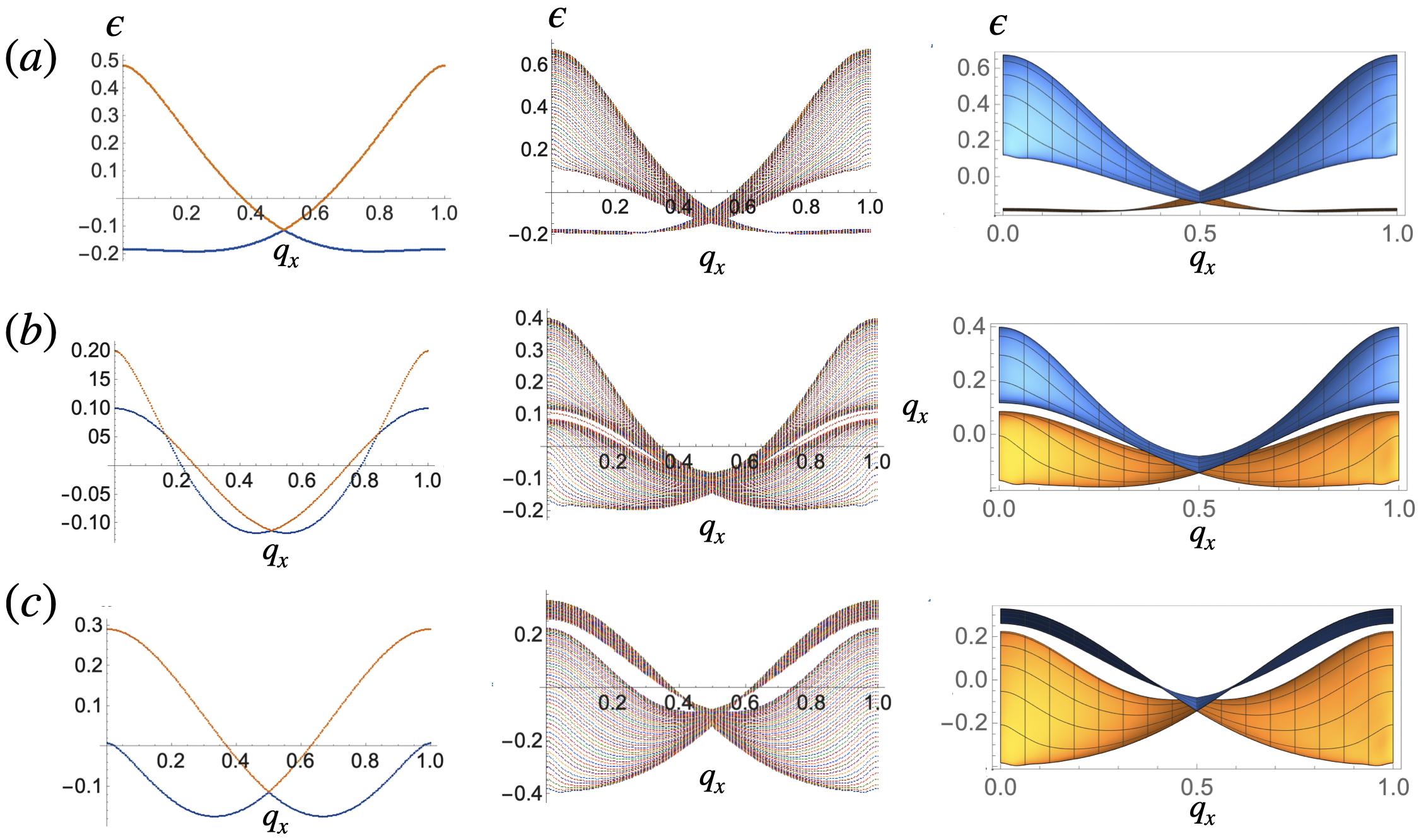}
\caption{Band spectrum of chains (left panel), stripes (middle panel) and lattices (right panel, side view at $q_z=0$) at $\mathcal{K}>\rm {\mathcal K}_c^\Lambda$,  $\kappa=\frac{\pi}{2}$ and (a) $\Lambda=0.5$, (b) $\Lambda=0.7$ and (c) $\Lambda=1$. Momenta along $\hat{x}$ and $\hat{z}$ are measured in units of $G_x$ and $G_z$ respectively. The minimum eigenvalues $\epsilon$ are in units of $\mathcal{D}$.}
\label{fig:f2}
\end{figure}
\subsection{Magnetic configurations that minimize Eq.~(\ref{eq:Energy}).} 
Next, we examine the effect of ${\mathcal K}$ and $\Lambda$ on the energy spectrum of the dipolar systems. In the paper, $\Lambda$ is tuned in the domain $0<\Lambda<3a_0$.
\subsubsection{Energetics of finite systems.} 
\label{sec:energetics}
Depending on $\Lambda$ and $\mathcal{K}$, finite modulated chains of point dipoles minimize Eq.~(\ref{eq:Energy}) in the magnetic configurations illustrated in Fig.~\ref{fig:f1}. For small anisotropy $\mathcal{K}<{\mathcal K}_c^{\Lambda}$, the collinear magnetic order, Figs.~\ref{fig:f1}(c-d) is favored. For $\mathcal{K}>{\mathcal K}_c^{\Lambda}$ the parallel magnetic states (Figs.~\ref{fig:f1}(a-b)) minimize energy. From Eq.~(\ref{eq:Energy}) one can find the critical anisotropy that sets the boundary between collinear and parallel configurations. For chains, this is found to be ${\mathcal K}_c^{\Lambda}=2\mathcal{D}\left(\frac{3 \zeta(3)}{4}-\frac{1}{2} \psi^{(2)}\left(\sqrt{\Lambda^2+1}\right)\right)$ where $\psi$ is the PolyGamma function and $\zeta$ the Riemann zeta function (details can be found in \cite{supp}). Thus $\rm {\mathcal K}_c^{\Lambda}$ becomes smaller as $\Lambda$ grows. For 2D lattices ($0<\Lambda<3a_0$), the critical anisotropy becomes $\rm {\mathcal K}_c^{\Lambda,L}\sim {\mathcal K}_c^{\Lambda}+\mathcal{D}$ . 

Consider the case of low anisotropy ($\mathcal{K}<{\mathcal K}_c^{\Lambda}$). At $\Lambda>0$, chains relax into two possible collinear magnetic configurations along $\hat{x}$. One, at low $\Lambda$ (roughly $\Lambda<1$) is the ferromagnetic state shown in Fig.~\ref{fig:f1}(c), the other at larger amplitudes is the collinear antiferromagnetic state shown in Fig.~\ref{fig:f1}(d). 
For high anisotropy, $\rm {\mathcal K}>{\mathcal K}_c$ and $\Lambda\lesssim 1$ the antiferromagnetic parallel configuration along $\hat{z}$ is favored, Fig.~\ref{fig:f1}(a). At intermediate amplitudes $1\lesssim\Lambda\lesssim 2$ dimers consisting of nearest neighbour ferromagnetic parallel dipoles arranged in an antiferromagnetic fashion `down-down-up-up' (and time reversal) along $\hat{z}$ are favored (Fig.~\ref{fig:f1}(b)). A second dimer configuration, `up-down-down-up' (and time-reversal) competes energetically with the first. For $\Lambda>2$ the two sublattices behave like independent chains with twice the lattice constant. The dimerized configurations in finite chains are not perfect. They contain domain walls or kinks with dipoles pointing in the wrong direction, which disturb the perfect dimer order \cite{cisternas2021stable}. A few kinks also occur in the antiferromagnetic parallel state. However, they are practically absent in the collinear configurations. Snapshots product of the energy minimization of finite chains can be found in \cite{supp}.

Minimization of the dipolar energy in 2D systems results in stable magnetic configurations that preserve the magnetic order of its building blocks, as illustrated in Fig.~\ref{fig:f1} and shown in the next section. 
Hereafter, we focus on systems at large anisotropy $\rm \mathcal{K}>{\mathcal K}_c^\Lambda$. Their energy scale is set by $\mathcal{D}$. 
\subsubsection{Band spectrum in terms of $\Lambda$.} 
\label{sec:ebands}
In units of $\mathcal{D}$, the interacting dipolar hamiltonian can be written
\begin{eqnarray}
\mathcal{H}^{\rm dd}&=&\sum_{i\neq j}{\mathcal M}_j^{\dagger}\hat{\mathcal{J}}_{ij}{\mathcal M}_i=\nonumber \\  
& &\sum_{i\neq j}{\mathcal M}_j^{\dagger}\left(\begin{array}{cc}J^{(11)}_{ij} & J^{(12)}_{ij} \\J^{(21)}_{ij} &J^{(22)}_{ij}\end{array}\right){\mathcal M}_i,
\label{eq:energy}
\end{eqnarray}
Here, ${\mathcal M}_{\jmath}^{\dagger}=(\hat{\bm m}_\jmath^{(1)}, \hat{\bm m}_\jmath^{(2)})$, where $\hat{\bm m}_\jmath^{(1)}$ and $\hat{\bf m}_\jmath^{(2)}$ denote the magnetic moment of the j-th dipole belonging to sublattice $(1)$ and the j-th dipole belonging to sublattice $(2)$ respectively, see Fig.~\ref{fig:f1} (a). $\hat{\mathcal{J}}_{ij}$ denotes the interaction matrix between dipoles $i$ and $j$. In the parallel configuration along $\hat{z}$, the interaction among all dipoles that belong to the same sublattice in a chain reads $\sum_{i\neq j}J^{(11)}_{ij}\equiv  J^{(11)}_z=J^{(22)}_z=\sum_{n=1}^{\infty}\frac{1}{8n^3}$ while for dipoles in different sublattices $\sum_{i\neq j}J^{(12)}_{ij}\equiv J^{(12)}_z=J^{(21)}_z=\sum_{n=1}^{\infty}\frac{-1+3\Lambda^2}{(n^2+\Lambda^2)^{3/2}}$. In momentum space ${{\mathcal M}}_{{\bm q}}=\frac{1}{\sqrt{n}}\sum_\jmath {\mathcal M}_\jmath e^{\imath {\bm q}\cdot{\bm r}_\jmath}$ and Eq.~(\ref{eq:energy}) becomes 
\begin{eqnarray}
{\mathcal H}^{\rm dd}&=&\sum_{\bm q}{\mathcal M}_{\bm q}^{\dagger}\hat{\mathcal{J}_{\bm q}}{\mathcal M}_{\bm q},
\label{eq:hdd}
\end{eqnarray}
where $\hat{\mathcal{J}_{\bm q}}=\left(\begin{array}{cc}J_{{\bm q},z}^{(11)}& J_{{\bm q},z}^{(12)} \\J_{{\bm q},z}^{(12)*}  &J_{{\bm q},z}^{(11)}\end{array}\right)$. Diagonalization of  the interaction matrix $\hat{\mathcal{J}}_{\bm q}$ for chains results in two branches with energy $\epsilon_1(q_x)=J_{q_x,z}^{(11)}+J_{q_x,z}^{(12)}$ and $\epsilon_2(q_x)=J_{q_x,z}^{(11)}-J_{q_x,z}^{(12)}$ with $J_{q_x,z}^{(11)}=-\frac{1}{8}s_1$, $J_{q_x,z}^{(12)}=s_3+\Lambda^2 s_4$  \cite{supp}
and 
\begin{align}
s_1&=&\Phi_3\left(e^{2 i q_x}\right), 
\nonumber
\\
s_3&=&e^{i q_x f(\Lambda)} \Phi \left(e^{i
   q_x},3,f(\Lambda)\right),
\nonumber\\
   s_4&=&e^{i q_x f(\Lambda)} \Phi \left(e^{i
   q_x},5,f(\Lambda)\right)\nonumber
   \end{align}
 where $f(\Lambda)=\sqrt{1+\Lambda^2}$  and $\Phi$ is the Polylogarithmic function.
Fig.~\ref{fig:f2} shows the energy bands of infinite (with periodic boundary conditions) chains (left panel), stripes made out of 60 infinite chains stacked along $\hat{z}$ (middle), and 2D lattices (right panel). In the figure,  momenta along $\hat{x}$ and $\hat{z}$  are shown in units of the primitive vectors of the first Brillouin zone (1BZ), $G_x=\frac{2\pi}{2a_0}$ and $G_z=\frac{2\pi}{3a_0}$. Bands are obtained from the exact diagonalization of interaction matrices $\hat{\mathcal{J}_{\bm q}}$ at several values of $\Lambda$. 
Qualitatively, the spectra of chains, stripes, and lattices look alike. The main features like the band crossing at $\frac{G_x}{2}$ are preserved in the three types of arrays and remain invariant as $\Lambda$ grows. Examination of the wavevectors of the minimum eigenvalues of the interaction matrix reveals that for $\Lambda\lesssim0.7$ the two branches in the energy spectrum of chains correspond to the antiferromagnetic and dimer configurations shown in Fig.~\ref{fig:f1}(a-b). At the $\Gamma$ point, the minimum energy corresponds to the antiferromagnetic mode. Increasing $\Lambda$ augments the relative interaction strength between dipoles in the same sublattice. Therefore, at the band crossing at $\Lambda=0.7$, the second dimerized configuration becomes energetically favorable compared to the antiferromagnetic state in the chains, and now each energy branch corresponds to  one of the two dimer modes. 

The middle panel of Fig.~\ref{fig:f2} shows that bands in stripes sort into bundles or groups of bands. These bundles, whose number equals the number of energy branches in the chains, resemble the bulk bands of the lattices shown in the right panel of Fig.~\ref{fig:f2}. 

Overall, increasing $\Lambda$ reduces rather drastically the energy gap at the $\Gamma$ point of the 1BZ. In addition, larger amplitudes amplify the relative interaction strength between dipoles belonging to different chains. This fact manifests as a qualitative change in the dispersive character of the bands, which become flattered as $\Lambda$ grows. 
\section{Magnon spectrum}
\label{sec:waves}
Next we examine the collective transverse excitations of the dipoles magnetization vector with respect to the parallel states found in Section \ref{sec:energetics}.  Neglecting damping, the dynamics of the magnetization vector of a dipole $\jmath$, belonging to sublattice $\alpha$, $\bm {\hat{m}}_j^{\alpha}(t)$ is described by the  Landau-Lifshitz equation \cite{lakshmanan2011fascinating,osokin2018spin,galkin2005collective,bondarenko2010collective,verba2012collective,lisenkov2014spin,lisenkov2016theoretical}
\begin{eqnarray}
\frac{d{\bm {\hat{m}}}_j^{\alpha}}{dt} =\gamma({\bf B}_{\rm eff,j}^{\alpha}\times {\bm {\hat{m}}}_j^{\alpha})
\label{eq:LL}
\end{eqnarray}
 where $\gamma$ is the modulus of the gyromagnetic ratio, and the effective magnetic field ${\bm B}_{\rm {eff,j}}^\alpha=-\frac{\partial {\mathcal H}^{dd}}{\partial \bm {\hat{m}}_{j}^\alpha}$
consists of the dipolar field created by other magnetic dipoles,
\begin{eqnarray}
{\bm B}_{\rm eff,j}^{\alpha}=-\mu_0\sum_{i,\beta} \hat{\mathcal{J}}_{ij}^{\alpha,\beta}\cdot{\bm{\hat{m}}_i^{\beta}},
\label{eq:eff}
\end{eqnarray}
\begin{figure*}
\includegraphics[width=\textwidth]{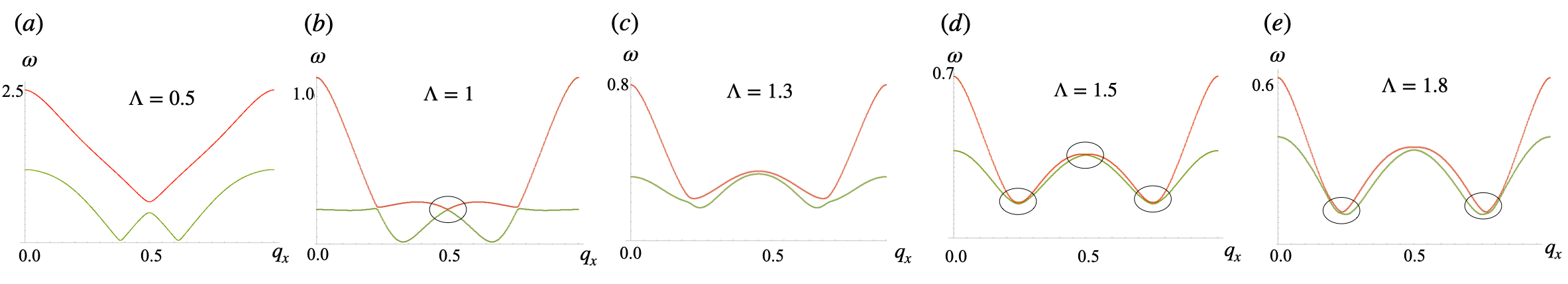}
\caption{Magnon spectrum of infinite chains (a) $\Lambda=0.5$, (b) $\Lambda=1$, (c) $\Lambda=1.3$, (d) $\Lambda=1.5$ and (e) $\Lambda=1.8$. Band touchings are encircled. In (d) note the band touching at the three encircled  points. Frequency $\omega$ is in units of $\gamma$ and $q_x$ is in units of $G_x$.}
\label{fig:f3}
\end{figure*}
\begin{figure}
\includegraphics[width=\columnwidth]{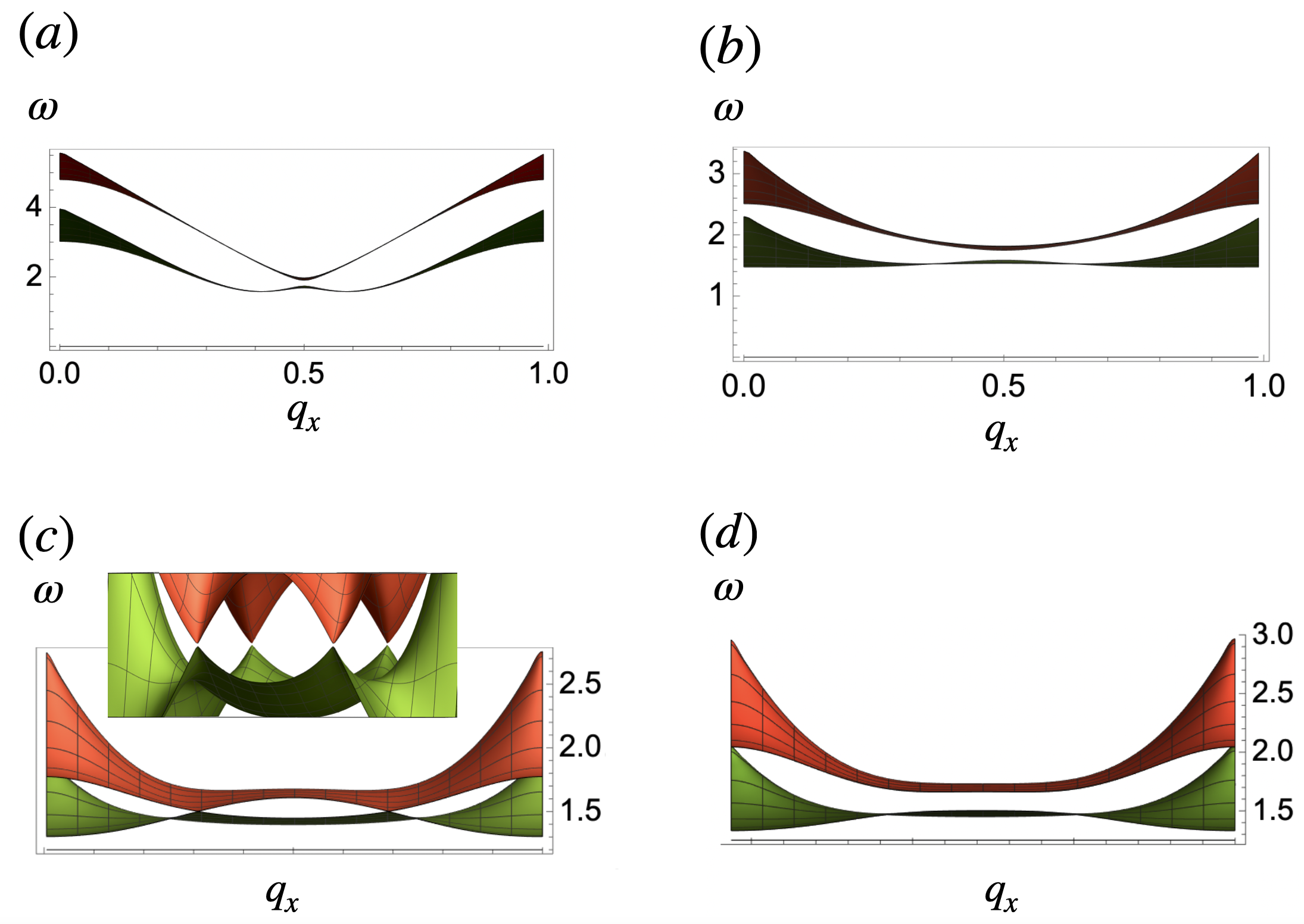}
\caption{Magnon spectrum of 2D lattices (side view at $q_z=0$) with  (a) $\Lambda=0.5$, (b) $\Lambda=1$, (c) $\Lambda=1.5$, and (d) $\Lambda=1.8$. In (c) two band touchings like Dirac points are located at $\left(\frac{G_x}{3},\frac{G_z}{2}\right)$ and $\left(\frac{2 G_x}{3},\frac{G_z}{2}\right)$ (and the two equivalent at $q_z=\frac{3G_z}{2}$) when $\Lambda=\frac{3}{2}$. They resemble the case of infinite chains.  For larger $\Lambda$ the group velocity of magnons decreases. Frequency $\omega$ is in units of $\gamma$ and $q_x$ is in units of $G_x$.}
\label{fig:f4}
\end{figure}
\begin{figure*}
\includegraphics[width=\textwidth]{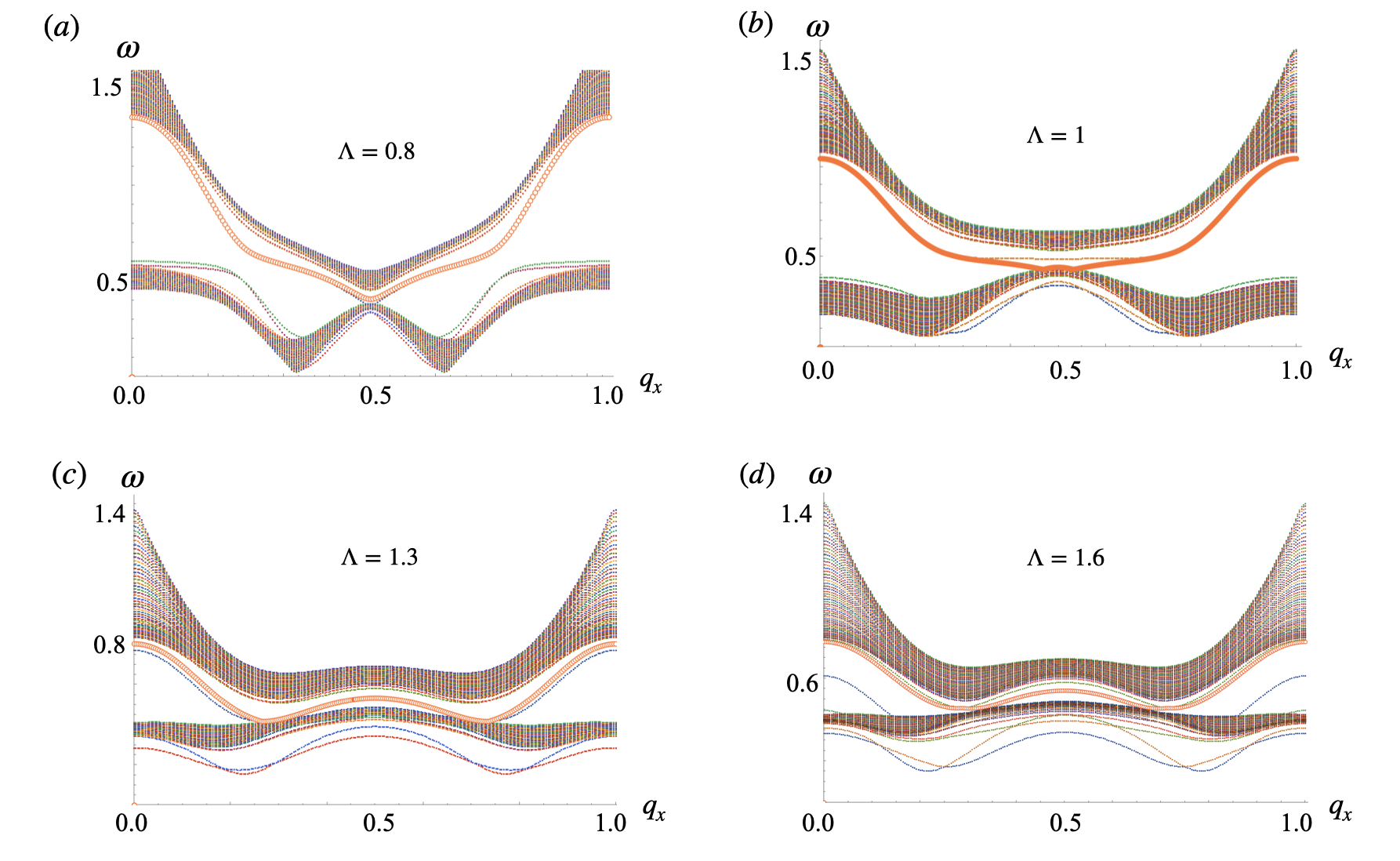}
\caption{Magnon spectrum of stripes at (a) $\Lambda=0.8$, (b) $\Lambda=1.0$, (c) $\Lambda=1.3$ and (d) $\Lambda=1.6$. Edge modes crossing the gap and joining the two bulk bands are highlighted.  Frequency $\omega$ is in units of $\gamma$ and $q_x$ is in units of $G_x$.}
\label{fig:f5}
\end{figure*}
where $\hat{\mathcal{J}}_{ij}^{\alpha,\beta}$ is the interaction matrix containing the geometrical aspects of the interactions between all dipoles in the array as discussed in the previous section. Hereafter we drop the hat from unit vectors.
Magnetization of the $j$th dipole in the stationary ground state ${\bm\mu}_j^{\alpha}=(0,0,\mu_{j}^{\alpha})$ is a unit vector in the direction of the static ground state. It points along a local axis which we call  $\bm{\hat{z}}$ and satisfies the system of equations:
\begin{eqnarray}
B_j^\alpha {\bm\mu}_j^\alpha= -\mu_0\sum_{i,\beta}\mathcal{\hat{J}}^{\alpha,\beta}_{z}\cdot{\bm\mu}_i^{\beta}
\label{eq:static}
\end{eqnarray}
 where $B_j^\alpha$ is the intrinsic scalar magnetic field acting on the $j$th dipole belonging to the $\alpha$ sublattice and $\mathcal{\hat{J}}^{\alpha,\beta}_{z}$ is the interaction matrix of the system in the stationary ground state. To find the dynamical equations describing small (linear) transverse
magnetization excitations, we use the following ansatz for the dipole magnetization:
\begin{eqnarray}
\bm{m}_j^{\alpha}(t)=({\bm\mu}_j^{\alpha}+{ \tilde{\bm{m}}}_j^{\alpha}(t))
\label{eq:linear}
\end{eqnarray}
where $\tilde{\bm {m}}_j^{\alpha}(t)=(\tilde{m}_{j,x}^{\alpha}(t),\tilde{m}_{j,y}^{\alpha}(t))$ is the small dimensionless deviation of the magnetization vector of the $j$th dipole from the static equilibrium state. Conservation of the length of the magnetization vector in each magnet requires that ${\bm\mu}_j\cdot \tilde{{\bm m}}_j=0$.
Eq.~(\ref{eq:linear}) in Eq.~(\ref{eq:LL}) yields:
\begin{eqnarray}
\frac{d{\tilde{\bm m}}_j^{\alpha}}{dt}=\gamma{\bm B}_{\rm eff,j}^{\alpha}\times\tilde{\bm {m}}_j^{\alpha}
\label{eq:linear2}
\end{eqnarray}
The effective field in terms of the scalar field and the stationary magnetization reads:
 \begin{eqnarray}
{\bm B}_{\rm eff,j}^{\alpha}=B_j^\alpha{\bm \mu}_j^\alpha-\mu_0\sum_{i,\beta} \hat{\mathcal{J}}^{\alpha,\beta}_{ij}\cdot\tilde{{\bm m}}_i^{\beta},
\label{eq:eff2}
\end{eqnarray}
which back into Eq.~(\ref{eq:linear2}) gives rise to 
\begin{equation}
\frac{\tilde{{\bm m}}_j^{\alpha}}{d t}=\gamma(B_j^\alpha{\bm\mu}_j^\alpha\times\tilde{{\bm m}}_j^{\alpha}+
\mu_0\sum_{i,\beta}{\bm\mu}_j^{\alpha}\times\hat{\mathcal J}^{\alpha,\beta}_{ij}\cdot\tilde{{\bm m}}_i^{\beta})
\label{eq:dm}
\end{equation}
Therefore equilibrium orientations of the uniform magnetization ${\bm\mu}_j^{\alpha}$ and internal fields $B_j^{\alpha}$ depend only on the index $\alpha$.  
At each sublattice, the linear spin-wave excitations have the form of plane waves. Thus Fourier transforming the magnetic excitation vector in time and space yields,
  \begin{eqnarray}
  \tilde{{\bm m}}_j^{\alpha}=\frac{1}{\sqrt{n}}\sum_q  \tilde{{\bm m}}^{\alpha}_{\bm{q}}e^{-(\imath{\bm q}\cdot{\bm r}_j^{\alpha}+\imath\omega t)},
\label{eq:fourier}
\end{eqnarray}
where ${\bm r}^{\alpha}_{j}$ is the vector position of a dipole belonging to sublattice $\alpha$ and located at the $j$th unit cell. Substituting Eq.~(\ref{eq:fourier})  back into Eq.~(\ref{eq:dm}) one obtains a finite dimensional eigenvalue problem:
\begin{eqnarray}
 -i\omega_{\bm q}\tilde{{\bm m}}_{\bm q}^{\alpha}=\gamma{\bm\mu}^{\alpha}_{\bm q}\times\sum_{\beta}(B^{\alpha}  \tilde{{\bm m}}_{\bm q}^{\alpha}+\mu_0\hat{\mathcal J}^{\alpha,\beta}_{\bm q}\cdot  \tilde{{\bm m}}^{\beta}_{\bm q})\nonumber,
\label{eq:eigen3}
\end{eqnarray}
  \begin{eqnarray}
 -i\omega_{\bm q}   \tilde{{\bm m}}_{\bm q}^{\alpha}=\gamma{\bm\mu}^{\alpha}_{\bm q}\times\sum_{\beta}\hat{\Omega}^{\alpha,\beta}_{\bm q}\cdot  \tilde{{\bm m}}^{\beta}_{\bm q},
\label{eq:eigen4}
\end{eqnarray}
where 
 \begin{eqnarray}
\hat{\Omega}^{\alpha,\beta}_{\bm q}=B^{\alpha}\hat{\delta}^{\alpha,\beta}+\mu_0\hat{\mathcal J}^{\alpha,\beta}_{\bm q},
\label{eq:omega}
\end{eqnarray}
 Since the vector $\tilde{{\bm m}}^{\beta}_{\bm q}=(\tilde{m}^{\beta}_{{\bm q},x}, \tilde{m}^{\beta}_{{\bm q},y})$ one can separate Eq.~(\ref{eq:eigen4}) along the $\hat{x}$ and $\hat{y}$ (transverse) directions: 
  \begin{eqnarray}
 -i\omega_{\bm q} \tilde{m}_{{\bm q},x}^{\alpha}=\gamma\left[{\bm\mu}^{\alpha}_{\bm q}\times\sum_{\beta}\hat{\Omega}^{\alpha,\beta}_{\bm q}\cdot\tilde{{\bm m}}^{\beta}_{\bm q}\right]_x,
\label{eq:eigenx}
\end{eqnarray}
  \begin{eqnarray}
  -i\omega_{\bm q} \tilde{m}_{{\bm q},y}^{\alpha}=\gamma\left[{\bm\mu}^{\alpha}_{\bm q}\times\sum_{\beta}\hat{\Omega}^{\alpha,\beta}_{\bm q}\cdot\tilde{{\bm m}}^{\beta}_{\bm q}\right]_y
\label{eq:eigeny}
\end{eqnarray}
with
  \begin{eqnarray}
\left[{\bm\mu}^{\alpha}_{\bm q}\times\sum_{\beta}\hat{\Omega}^{\alpha,\beta}_{\bm q}\cdot\tilde{{\bm m}}^{\beta}_{\bm q}\right]_x=-\mu^{\alpha}_{\bm q}\sum_\beta(B^\alpha\hat{\delta}^{\alpha,\beta}+\nonumber \\  \hat{\mathcal{J}}_{{\bm q},y}^{\alpha,\beta})\tilde{m}_{{\bm q},y}^{\alpha}
\label{eq:eigen2x}
\end{eqnarray}
and 
  \begin{eqnarray}
\left[{\bm\mu}^{\alpha}_{\bm q}\times\sum_{\beta}\hat{\Omega}^{\alpha,\beta}_{q}\cdot\tilde{{\bm m}}^{\beta}_q\right]_y=
\mu^{\alpha}_{\bm q}\sum_\beta(B^\alpha\hat{\delta}^{\alpha,\beta}+\nonumber \\  \hat{\mathcal{J}}_{{\bm q},x}^{\alpha,\beta})\tilde{m}_{{\bm q},x}^{\alpha}
\label{eq:eigen2y}
\end{eqnarray}
\subsubsection{Reconfigurable magnon frequencies.}
In terms of magnon creation $m_+=m_x+ i  m_y$ and annihilation $m_-=m_x- i m_y$ fields, the matrix form of the equations of motion Eq.~(\ref{eq:eigen2x}) and Eq.~(\ref{eq:eigen2y}) becomes 
  \begin{eqnarray}
\omega \hat{\sigma_z}\left(\begin{array}{c}m_+ \\m_-\end{array}\right)=\left(\begin{array}{cc}\hat{A} & \hat{B} \\\hat{B} & \hat{A}\end{array}\right)\left(\begin{array}{c}m_+ \\m_-\end{array}\right)
\label{eq:magnon}
\end{eqnarray}
where the right hand side matrix constitutes the magnon hamiltonian. The Pauli matrix  $\hat{\sigma}_z$, takes $+ 1$  for the creation field or particle space and $- 1$ for the annihilation field or hole space.
The $2\times 2$ matrices $\hat{A}=\left(\begin{array}{cc}a_1 & a_2 \\a_2^* & a_1\end{array}\right)$ and $\hat{B}=\left(\begin{array}{cc}b_1 & b_2 \\b_2^* & b_1\end{array}\right)$ with $a_1=\gamma \epsilon_0+\frac{\gamma}{2}(J_{{\bm q},x}^{(11)}+J_{{\bm q},y}^{(11)})$, $a_2=\gamma \epsilon_0+\frac{\gamma}{2}(J_{{\bm q},x}^{(12)}+J_{{\bm q},y}^{(12)})$, $b_1=\frac{\gamma}{2}(J_{{\bm q},x}^{(11)}-J_{{\bm {\bm q}},y}^{(11)})$, $b_2=\frac{\gamma}{2}(J_{{\bm q},x}^{(12)}-J_{{\bm q},y}^{(12)})$ and $\epsilon_0$ is the energy of the stationary magnetic state (hereafter we set $\epsilon_0=0$). Expressions for $J_{{\bm q},x}^{11}$ and $J_{{\bm q},y}^{12}$ are shown in \cite{supp}. The magnon hamiltonian can be written as:
\begin{equation}
\mathcal{H}=\hat{\sigma}_z\otimes(\hat{t}_0+\hat{t}_1)+\hat{\sigma}_x\otimes(\hat{t}_2+\hat{t}_3)
\label{eq:hmag}
\end{equation}
where $\hat{\sigma}_j$ is the jth Pauli matrix, $\otimes$ denotes Kronecker product, $\hat{t}_0=\left(\begin{array}{cc}a_1 & 0 \\0 & a_1\end{array}\right)$, $\hat{t}_1=\left(\begin{array}{cc}0 & a_2 \\a_2^* & 0\end{array}\right)$, $\hat{t}_2=\left(\begin{array}{cc}b_1 & 0 \\0 & b_1\end{array}\right)$ and $\hat{t}_3=\left(\begin{array}{cc}0 & b_2 \\b_2^* & 0\end{array}\right)$.
In Eq.~(\ref{eq:hmag}) the term multiplying $\hat{\sigma}_z$ is a mass term responsible for the gap,  while the term multiplying $\hat{\sigma}_x$ is proportional to the group velocity of the spin waves or magnon speed.
Eigenfrequencies for collective spin wave modes in the particle space can be written as
\begin{equation}
\omega_{1,2}^2=\frac{\gamma^2}{4}(a_1^2 \pm 2a_1a_2+a_
2^2+b_1^2 \pm 2b_1b_2+b_2^2)
\label{eq:freq1}
\end{equation}
where $a_1$, $a_2$, $b_1$ and $b_2$ can be written in terms of $s_1$, $s_3$ and $s_4$ \cite{supp}.
Figs.~\ref{fig:f3},~\ref{fig:f4} and ~\ref{fig:f5} show the effect of $\Lambda$ in the magnon spectrum of chains, lattices and stripes respectively. They result from the exact diagonalization of Eq.~(\ref{eq:hmag}). Similar to the stationary energy spectrum, the frequency dispersion in 2D systems resembles magnon bands in chains. Smaller values of $\Lambda$ allow for a wider range of frequencies for magnon excitation at the $\Gamma$ point in all the arrays.  

$\Lambda$ has a strong effect on the group velocity of the spin waves: larger amplitudes yield flatter bands and therefore lower magnon speeds around the middle of the spectrum. Compare, for instance, frequency slopes of systems with $\Lambda=0.5$ and $\Lambda=1.8$ in Figs.~\ref{fig:f4}(a) and (d) respectively.
Notable features are the band touchings in chains and lattices. In 1D, the first touching occurs in the middle of the spectrum $q_0=\frac{G_x}{2}$ at $\Lambda=1\equiv\Lambda_1$, Fig.~\ref{fig:f3}(b). At $\Lambda>\Lambda_1$ the gap opens again until $\Lambda=1.5\equiv \Lambda_2$ where the touching at $q_0$ reappears and two additional band touchings arise at $q_1\sim\frac{G_x}{4}$ and $q_2\sim 3\frac{G_x}{4}$, Fig.~\ref{fig:f3}(d). For $\Lambda>\Lambda_2$ the gap at $q_1$ and $q_2$ remains closed while that at $\frac{G_x}{2}$ reopens. One can estimate the critical amplitude for which the magnonic gap closes by equating the eigenfrequencies $\omega_1$ and $\omega_2$ in Eq.~(\ref{eq:freq1}). In 1D, for $q_x=\frac{G_x}{2}$ this yields the equality
\begin{equation}
\Lambda^2_{q_x=\frac{\pi}{2}}=\frac{1}{2}\frac{\Phi \left(\imath,3,\sqrt{1+\Lambda^2}\right)}{\Phi \left(\imath,5,\sqrt{1+\Lambda^2}\right)}
\end{equation}
which is satisfied at $\Lambda\sim \Lambda_1$.

In the case of lattices the band touching happens at $\Lambda_2$ at two points of the 1BZ, $\left(\frac{G_x}{3},\frac{G_z}{2}\right)$, $\left(2\frac{G_x}{3},\frac{G_z}{2}\right)$ as shown in Fig.~\ref{fig:f4}(c). Here the inset shows a close up of the band spectrum about the band touching points (and the equivalent $\left(\frac{G_x}{3},3\frac{G_z}{2}\right)$ and $\left(2\frac{G_x}{3},3\frac{G_z}{2}\right)$ ) which resembles Dirac cones. At all other $\Lambda$ the spectra of the lattices are gapped.

In stripes, Fig.~\ref{fig:f5} shows magnon bands that cross the gap between the two bulk bands (highlighted in orange). Isolated bands located above and below the bulk bands arise as well for large $\Lambda$, as the ones shown below the lowest frequency band in Fig.~\ref{fig:f5}(d). 

 \section{Topological Bands and Edge Modes.}
 \label{sec:chern}
In previous sections, we showed that the spin-wave spectrum acquires forbidden frequency band gaps due to the periodic modulation of the dipolar arrays. In this section, we revisit the spin-wave spectrum of the stripes. We note in Fig.~\ref{fig:f5} the onset of features in between the two bulk bands, which resemble edge modes. Aimed to unveil the nature of the in-gap modes of Fig.~\ref{fig:f5} we computed the Chern number of the volume bands associated with them. A non-zero Chern integer, $c_n$, for spin-wave volume bands results in the emergence of chiral spin-wave edge modes. These topological edge bands have chiral dispersion that favors the unidirectional propagation of magnetic degrees of freedom for a frequency between the gap. In addition, they are robust to intrinsic and externally induced disorder \cite{girvin2019modern}.  
 It is expected that finite Chern integers, $c_n$, result from strong spin-orbit coupled interactions \cite{shindou2013chiral,peter2015topological}. With an inner product between magnetization and position vectors, the magnetic dipolar interaction locks the relative rotational angle between the spin space and orbital space, similar to what the relativistic spin-orbit interaction does in electron systems \cite{shindou2013chiral}. As a result of the spin-orbit locking, the complex-valued character in the spin space is transferred into wave functions in the orbital space.
\subsection{Gauge connection and the associated Berry curvature in the lattice.}
 \label{sec:con}
The sense of motion and the number of chiral modes in a system is determined from the magnitude and sign of the topological number for volume mode bands below the bandgap. $c_n$, can be changed only by closing the gap \cite{girvin2019modern,shindou2013chiral}. 
\begin{table}[bt]
\begin{tabular}{|c|c|c|c|c|c|c|c|c|c|c|}
\hline
$\Lambda$&0.2&0.5&0.7&1&1.2&1.4&1.5&1.6&1.8&2.0\\
\hline
$c_1$&1&1&1&1&1&1&1&-1&-1&-1\\ 
\hline
$c_2$&-1&-1&-1&-1&-1&-1&-1&1&1&1\\ 
\hline
\end{tabular}
\caption{Chern numbers, of the magnon volume bands in 2D lattices at several values of $\Lambda$. Note the exchange of the Chern number between the two bands at $\Lambda_2=1.5$.}
\label{table:1}
\end{table}
\begin{figure}
\includegraphics[width=\columnwidth]{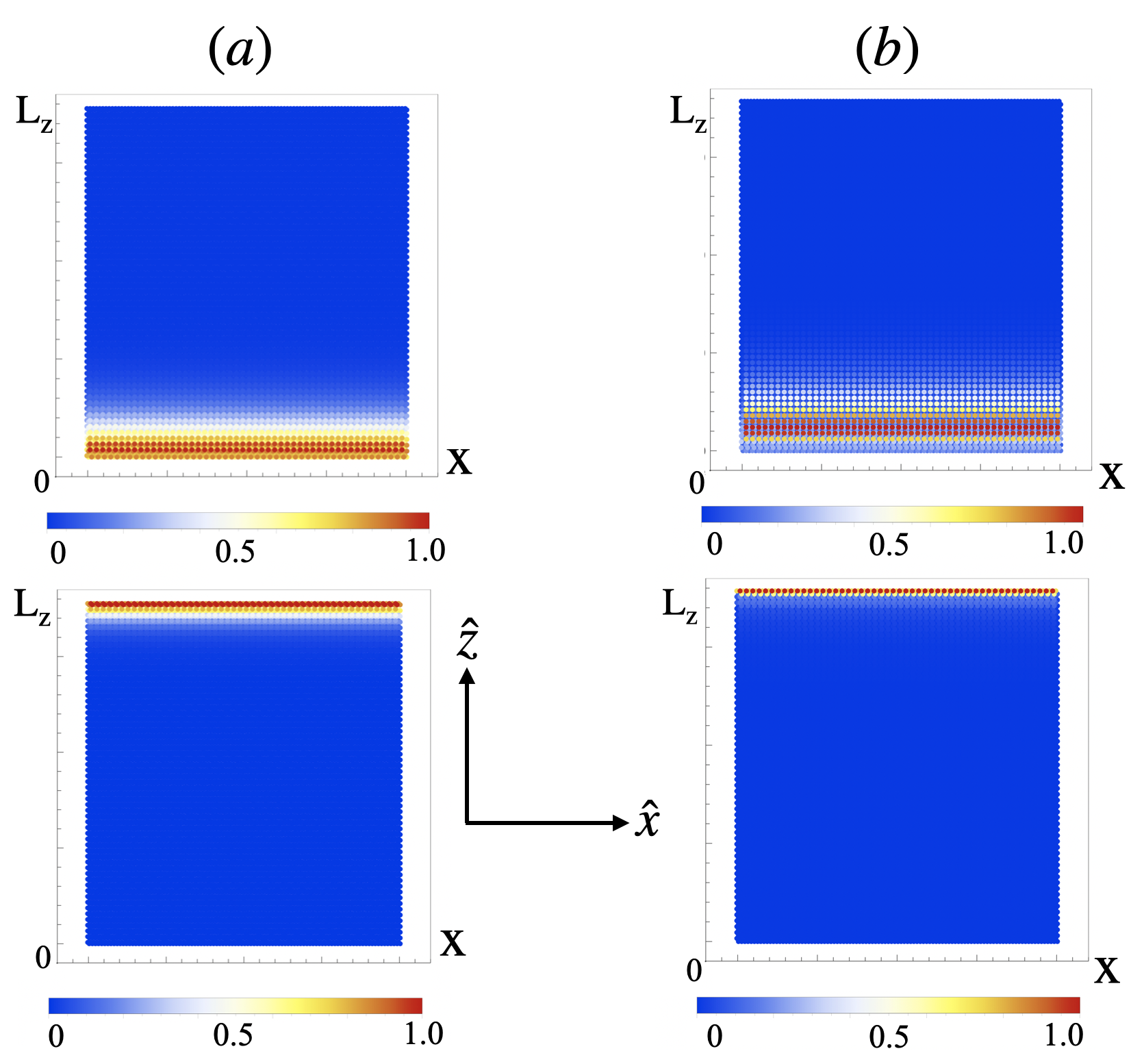}
\caption{Color map with the amplitude of the magnon Bloch wave function of the in gap eigenfrequencies of Figs.~\ref{fig:f5}(a) and (d) at each position of the stripes. (stripes with $L_z=60$ rows along the $\hat{z}$ axis) at (a) $\Lambda=0.8$ and (b) $\Lambda=1.6$. In both cases, edges states localized at $z=0$ and $z=L_z$ are apparent. The color code, shown underneath each figure spans from 0 in blue to 1 in red.}
\label{fig:f6}
\end{figure}
\begin{figure}
\includegraphics[width=\columnwidth]{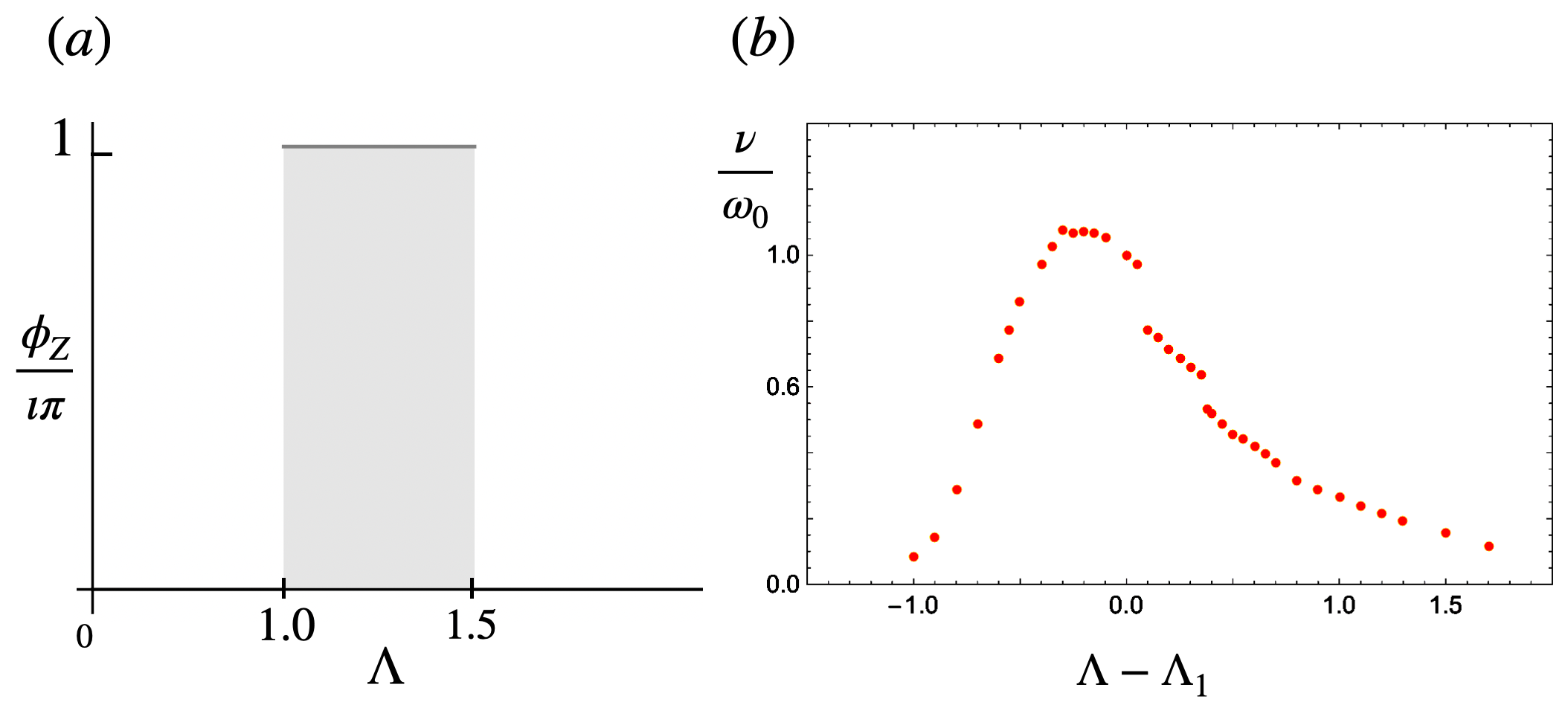}
\caption{(a) Zak phase of infinite chains. Topological phase transitions occur at $\Lambda_1=1.0$ and $\Lambda_2=1.5$.  (b) The group velocity of magnonic edge states $\nu$ depicts jump discontinuities at $\Lambda_1$ and $\Lambda_2$.}
\label{fig:f7}
\end{figure}
\begin{figure}
\includegraphics[width=\columnwidth]{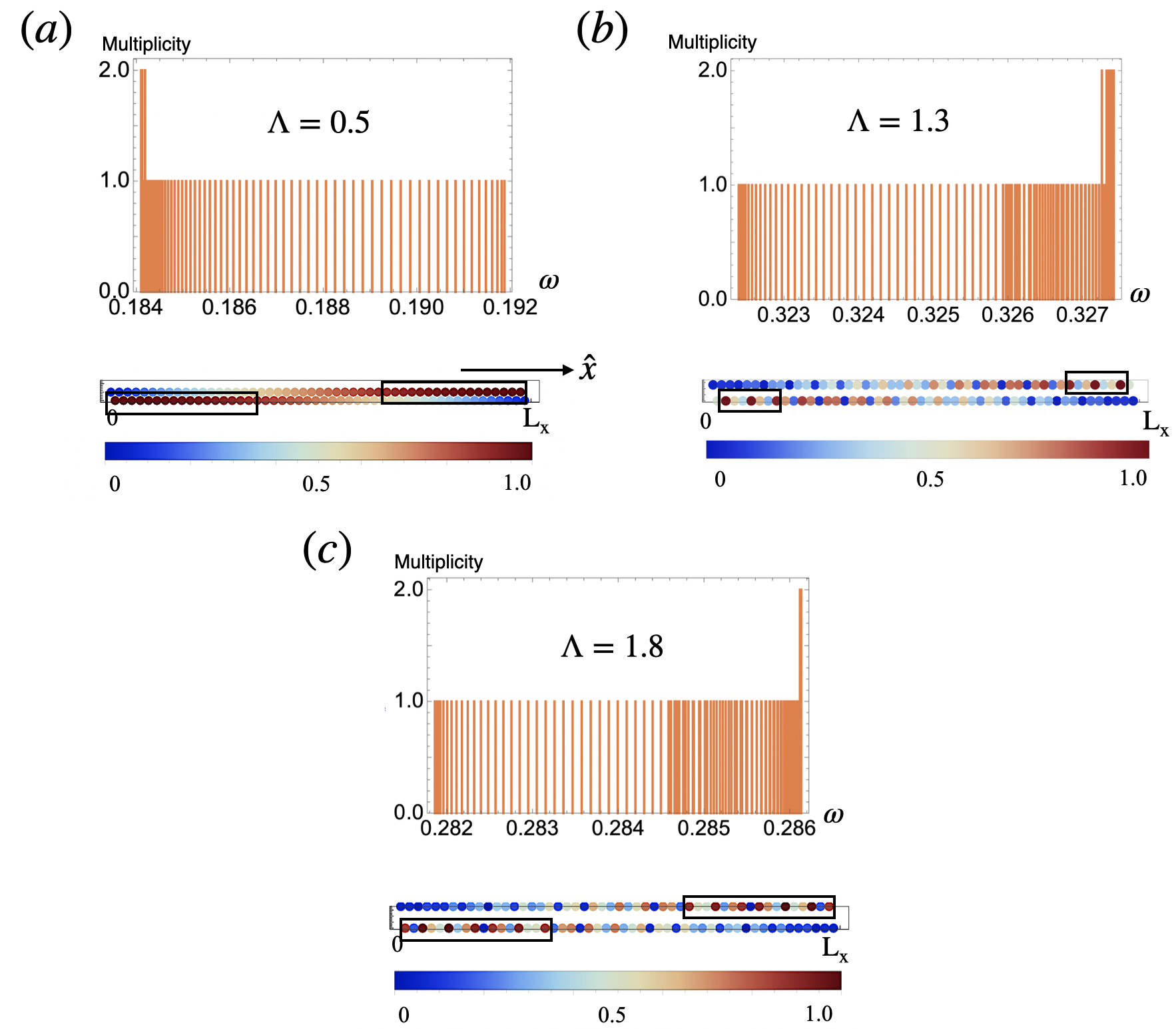}
\caption{Histograms show the multiplicities of the magnon eigenfrequencies $\omega$ of open chains with $n=100$ dipoles. Magnetic stable configurations have the largest frequency in these plots. The color maps underneath each histogram show the respective amplitude of the magnon Bloch wave function of the largest eigenfrequency as a function of the position in the chain ($L_x$ denotes the length of the chains). The color code, shown underneath each figure spans from 0 in blue to 1 in red. Figures correspond to chains with (a) $\Lambda=0.5$, (b) $\Lambda=1.3$ and (c) $\Lambda=1.8$. The localization of the wave function at the edges is apparent in (b).}
\label{fig:f8}
\end{figure}
\begin{figure}
\includegraphics[width=\columnwidth]{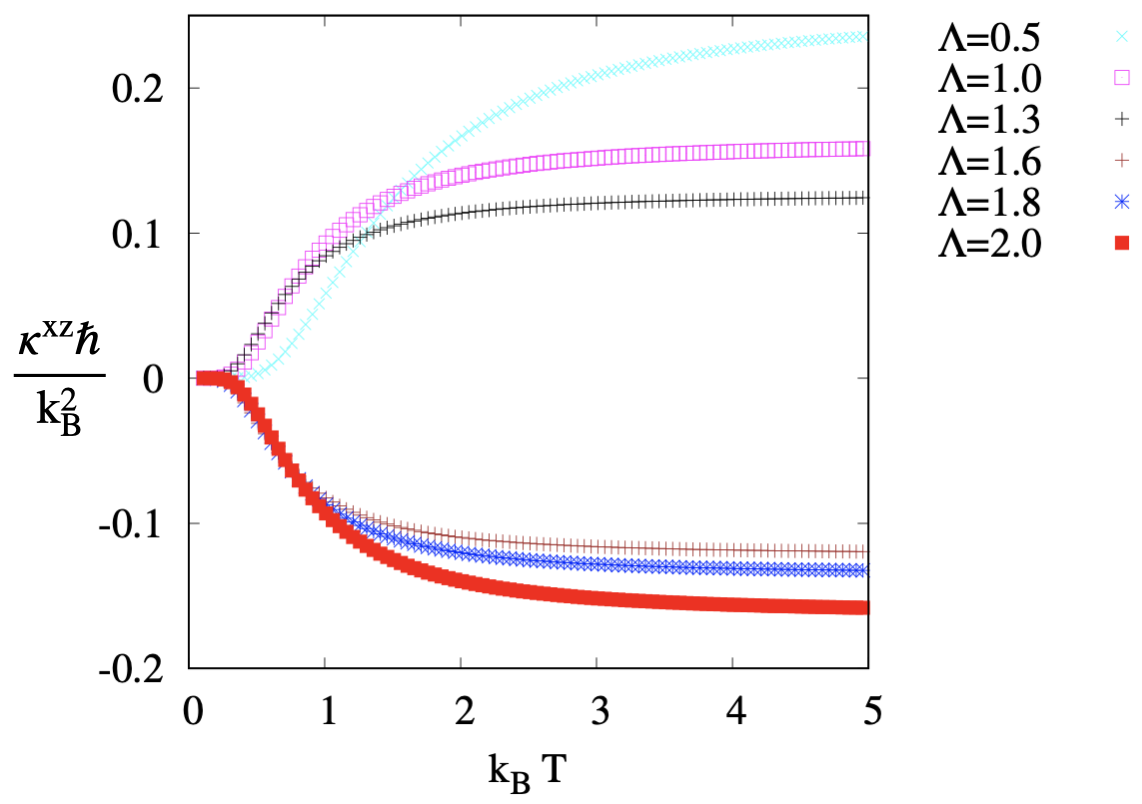}
\caption{Thermal conductivity as a function of $k_BT$ at different $\Lambda$. Note the change in the sign of $\kappa^{xz}$ for $\Lambda>1.5$.}
\label{fig:f9}
\end{figure}
Here, frequency bands are computed on a discretized Brillouin zone. Thus, we follow the approach of reference \cite{fukui2005chern} to compute the Berry phase \cite{zak1989berry} and the Chern integer using wave functions given on such discrete points.
The Chern number assigned to the $n$th band is the integral of fictitious magnetic fields: that is, field strengths of the Berry connection. It is defined by
  \begin{eqnarray}
  c_n=\frac{1}{2\pi\imath}\int_{\tau^{2}} d^2 qF_{xz}(q)
 \label{eq:chern}
\end{eqnarray}
where $\tau^2$ denotes the Brillouin zone torus.
The gauge connection (gauge field) $A_\mu(q)$ ($\mu=x,z$) and the associated field strength $F_{xz}(q)$ or Berry curvature are given by  
  \begin{eqnarray}
A_\mu&=&\langle n(q)|\partial_\mu| n(q) \rangle \\\nonumber
F_{xz}&=&\partial_xA_z(q)-\partial_zA_x(q)
 \label{eq:berry}
\end{eqnarray}
where $| n(q) \rangle$ is the normalized wave function of the $n$th (particle) Bloch band such that $H(q)| n(q) \rangle=\omega_n(q)| n(q) \rangle$.  
On the discrete Brillouin zone with lattice points $q_l$, $l=1,...n_xn_z$ the lattice field strength is given by 
  \begin{align}
\tilde{F}_{xz}(q_l)\equiv\ln U_x(q_l)U_z(q_l+\hat{x})U_x(q_l+\hat{z})^{-1}U_z(q_l)^{-1}\\\nonumber
-\pi<\frac{1}{i}\tilde{F}_{xz}(q_l)\leq \pi \nonumber
 \label{eq:discberry}
\end{align}
where $U_\mu(q_l)$ is a U(1) link variable from a $n$th band,
  \begin{eqnarray}
U_\mu(q_l)\equiv \frac{\langle n(q_l)| n(q_l+\hat{\mu}) \rangle}{|\langle n(q_l)| n(q_l+\hat{\mu}) \rangle|}
 \label{eq:link}
\end{eqnarray}
and the gauge invariant Chern number on the lattice \cite{fukui2005chern} associated to the $n$th band is finally
  \begin{eqnarray}
 \tilde{c}_n\equiv\frac{1}{2\pi i}\sum_l \tilde{F}_{xz}(q_l)
 \label{eq:chern2}
\end{eqnarray}
\subsection{Topological magnons at the edges of stripes}
\label{sec:edges}
Using the previous approach, we computed the Chern numbers for lattices corresponding to the bulk bands of the stripes shown in Fig.~\ref{fig:f5}. The results are presented in Table~\ref{table:1}. For $\Lambda\leq\Lambda_2$ the lowest frequency band has Chern number $c_1=1$ while the upper band has $c_2=-1$. This topological phase of the magnon dispersion is characterized in terms of $c_1$ and $c_2$ and denoted (1,-1). At $\Lambda>\Lambda_2$ the Chern numbers of the two bands are exchanged, the topological phase is denoted (-1,1).
Going back to Fig.~\ref{fig:f4}(c), we note two band touchings for the case $\Lambda=\Lambda_2$, at points $(\frac{G_x}{3},\frac{G_z}{2})$ and $(\frac{2G_x}{3},\frac{G_z}{2})$ of the 1BZ. There, the bands form approximated gapless Dirac spectra (inset in Fig.~\ref{fig:f4}(c)). 
A band touching point in the 3D parameter space $(q_x,q_z,\Lambda)$ plays the role of a dual magnetic monopole. The corresponding dual magnetic field is a rotation of the three component gauge field $\bm A_n=(A_{nx},A_{nz},A_{n,\Lambda})$, $\bm{B}_n=\bm\nabla\times \bm A_n$, where $\bm\nabla\equiv (\partial_{qx},\partial_{qz},\partial_\Lambda)$ and $n$ specifies either of the magnonic bands which form the band touching. Following Eq.~(\ref{eq:berry}), $A_{n,\Lambda}=\langle n(q)|\partial_\Lambda | n(q) \rangle$. At the band touching point, the dual magnetic field for the respective bands has a dual magnetic charge, whose strength is quantized to be $2\pi$ times an integer \cite{girvin2019modern}. Because the Chern integer $c_n$ can be regarded as the total dual magnetic flux penetrating through the constant $\Lambda$ plane, the Gauss theorem implies that when $\Lambda$ goes across the $\Lambda=\Lambda_2$ plane, the Chern integer for the lowest magnonic band $c_1$ changes by unit per each touching point. Hence, due to the two band touchings $c_1|_{\Lambda>\Lambda_2}-c_1|_{\Lambda<\Lambda_2}=2$, which explains the exchange of Chern numbers between bands at $\Lambda_2$.  
According to the bulk-edge correspondence, principle \cite{girvin2019modern} the number of in gap one-way edge states is determined by the winding number of a given band $n$, that is the sum of all the Chern numbers of the band up to band $n$. Consequently, stripes should realize one topological edge mode at each edge. Figs.~\ref{fig:f6} (a) and (b), show the amplitude of the magnon Bloch wave function for the modes crossing the band's gap in stripes made out of 60 chains, at $\Lambda=0.8$ and $\Lambda=1.6$ respectively. The localization of the in-gap modes at the edges of the stripes is apparent in both cases. 
\subsubsection{Thermomagnetic Hall transport}
\label{sec:hall}
Upon applying a temperature gradient, the magnon Hall effect MHE allows a transverse heat current mediated by magnons in two dimensions. The MHE was discovered in the ferromagnetic insulator $\rm Lu_2V_2O_7$ \cite{onose2010observation} and explained in terms of uncompensated magnon edge currents in two dimensions \cite{matsumoto2011theoretical,matsumoto2011rotational}. The relevant quantity characterizing the MHE is the thermal Hall conductivity. Similar to electronic systems, the thermal Hall conductivity is related to the Berry curvature of the eigenstates. The intrinsic contribution to the transverse thermal conductivity
\begin{equation}
\label{hall }
\kappa^{xz}=-\frac{k_B^2 T}{4\pi^2\hbar}\sum_i\int_{BZ}\theta(\rho_i)F^{xz}_i(\bm q)\bm dq
\end{equation}
is intimately related to the Chern numbers defined in Eq.~(\ref{eq:chern}). The sum is over all bands $i$ in the magnon dispersion, and the integral is over the 1BZ. $\rho_i$ is the Bose distribution function and the function,
\begin{equation}
\label{theta}
\theta(x)\equiv (1+x)(\ln\frac{1+x}{x})^2-(\ln x)^2
-2Li_2(-x)
\end{equation}
where $Li_2$ is the dilogarithm, $T$ the temperature, and $k_B$ is the Boltzmann constant. The thermal Hall conductivity can be interpreted as the Berry curvature weighed by the $\theta$ function \cite{matsumoto2011rotational}. The sign of $\kappa^{xz}$  
depends on the topological phase of the bulk system. This dependence can be understood in terms of edge modes and their propagation direction \cite{mook2014edge}. The topological phases (1,-1) and (-1,1) produce one edge mode. They differ in the slope of their dispersion. In the first case, the nontrivial edge mode propagates to the right, while in the second, it does to the left. The sign of $\kappa^{xz}$, and therefore the direction of the heat transport in a given topological phase, depends on the occupation probability of the edge magnons. When there is more than one edge mode in the same phase with different slopes, two propagation directions are possible depending on $T$. Since $\kappa^{xz}$ is weighted by the function $\theta$, edge modes propagating in different directions may induce cancellation of the transverse thermal conductivity at high energies. If all nontrivial edge modes propagate in the same direction, the sign of the thermal Hall conductivity is fixed within the topological phase, and its sign does not depend on temperature. Here,  phase (1,-1) has $\kappa^{xz}>0$, while in phase (-1,1) $\kappa^{xz}<0$. Fig.~\ref{fig:f9} shows the thermal Hall conductivity in terms of temperature for several values of $\Lambda$. As expected, $\kappa^{xz}$ has opposite sign for lattices with $\Lambda<1.5$ and lattices with $\Lambda>1.5$.

\subsection{Zak phase in infinite chains and topological magnons in open chains.}
\label{sec:zak}
The strong resemblance between the frequency spectrum of 1D and 2D systems suggests that modulated chains could manifest topological behavior \cite{li2021magnonic}. To examine this possibility, we have computed the Berry phase of 1D systems over a non-contractible loop $\ell$ of the 1BZ. This invariant is known as the Zak phase \cite{atala2013direct, zak1989berry} and is defined as: 
\begin{eqnarray}
Z=i\int_\ell{\bm dq}\cdot {\bm A}_n(\bm{q})=i \int_{-G_x/2}^{G_x/2} {\bm dq}\cdot {\bm A}_n(\bm{q})
 \label{eq:zak1}
\end{eqnarray}

When the 1D system has inversion symmetry a non zero Zak phase indicates that the system is in a topological phase \cite{zhang2013topological}. It is easy to verify, that the magnon hamiltonian in Eq.~(\ref{eq:hmag}) has an inversion symmetry $\mathcal{I}\mathcal{H}(q_x)\mathcal{I}^{-1}={H}(-q_x)$, with respect to the unitary operators $\mathcal{I}=\sigma_0\otimes\sigma_x$ and $\mathcal{I}=\sigma_x\otimes\sigma_x$, with $\sigma_0$ the identity matrix.

Following the approach of Section~\ref{sec:con} the gauge field  \cite{fukui2005chern} in the 1D lattice reads,
\begin{eqnarray}
\tilde{A}_x(q_l)&=\ln U_x(q_l)\\
\nonumber
\tilde{Z}_n&=i\sum_l\tilde{A}_x(q_l)&=i\sum_l\ln U_x(q_l)
\nonumber
\end{eqnarray}
Fig.~\ref{fig:f7}(a) shows the result of the Zak number of infinite modulated chains in terms of $\Lambda$. For modulation amplitudes in the intervals $0<\Lambda<\Lambda_1$ and $\Lambda>\Lambda_2$ the Zak phase is zero, while for $\Lambda_1<\Lambda<\Lambda_2$ the phase is equal to $\pi$ and the system is in a topological phase. There is a link between this bulk topological invariant, and the presence of topologically protected end states \cite{girvin2019modern}. Fig.~\ref{fig:f8} shows a color map of the magnon wave function amplitude with largest eigenfrequency in chains with $n=100$ dipoles at different values of $\Lambda$. The histogram in each figure shows the multiplicities of the magnon eigenfrequencies of such chains. 
In the interval $\Lambda_1<\Lambda<\Lambda_2$, Fig.~\ref{fig:f8}(b) the density map shows well-localized end states. However for $\Lambda<\Lambda_1$ and $\Lambda>\Lambda_2$, Figs.~\ref{fig:f8}(a) and (c) the magnon mode is delocalized and distributed all over the chain. Consequently we conclude that the chain realizes protected topological end states in the interval $\Lambda_1<\Lambda<\Lambda_2$ \cite{wang2018topological}.

Going back to Fig.~\ref{fig:f3}, we note that the quantized jumps in the Zak phase shown in Fig.~\ref{fig:f7}(a) coincide with the two band touchings at $\Lambda_1$ and $\Lambda_2$, shown respectively in Fig.~\ref{fig:f3}(b) and Fig.~\ref{fig:f3}(d). The previous analysis of the energy spectrum showed that at $\Lambda\sim 0.7$, dipoles modify their magnetic equilibrium state from an antiferromagnetic to a dimer configuration (Fig.~\ref{fig:f1}(b)). Thus, the band touching in the magnon spectrum and the change in the Zak phase at $\Lambda_1$ can be attributed to this fact. Furthermore, the second jump of the Zak number and simultaneous band touchings at $\Lambda_2$ coincides with the magnetic transit from the `down-down-up-up' to the `up-down-dow-up' dimerized configurations \cite{marti2021absence}.
\subsection{Effective model near band touching points.}
\label{sec:effective}
In 1D the band touching at $p_1=(q_0,\Lambda_1)$ (Fig.~\ref{fig:f3}(b)) can be seen as a single Dirac point around which the frequency dispersion for both bands can be approximated by a linear function 
\begin{eqnarray}
\omega^{(p_1)}_{1,2}\sim \pm\nu|q|\nonumber
\end{eqnarray}  
with $\nu$ the speed of magnons in the chain. The singular structure of the frequency dispersion near the band touching can be studied using degenerate perturbation theory \cite{shindou2013chiral}. For the magnon hamiltonian studied above, it takes the form 
\begin{eqnarray}
H_p=H_{1}+V_p\nonumber
\end{eqnarray} 
with $H_1=\mathcal{H}(p_1)$ and $V_p=H_p-H_{1}$. At the touching point, $H_{1}$ has twofold degenerate eigenstates $| d_j \rangle$ (j=1,2) with eigenfrequency $\omega_0$ $(>0)$, that satisfies $H_{1}| d_j \rangle=\omega_0\sigma_z | d_j \rangle$ \cite{supp}. On introducing the perturbation $V_p$, the degeneracy is split into two frequency levels. The eigenstate for the respective eigenfrequency is determined on the zero order of $p-p_1$ as
\begin{eqnarray}
T_p=T_1U_p+\mathcal{O}(|p-p_1|)\nonumber
\end{eqnarray}   
where the matrix $T_1$ diagonalizes $H_1$ and the unitary matrix $U_p$ diagonalizes a 2 by 2 hamiltonian $h_{p}$ formed by the twofold degenerate eigenstates,
\begin{eqnarray}
h_p=\left(\begin{array}{cc}d_1^\dagger V_p d_1 & d_1^\dagger V_p d_2 \\d_2^\dagger V_p d_1 & d_2^\dagger V_p d_2\end{array}\right)\nonumber
\end{eqnarray}  
 In Fourier space this can be written as 
\begin{eqnarray}
h_q= \left(\begin{array}{cc}f_1(q,\Lambda) & f_2(q,\Lambda) \\f_3(q,\Lambda) & f_4(q,\Lambda)\end{array}\right)\nonumber
\end{eqnarray}  
where we find that $f_1(q,\Lambda)=-f_4(q,\Lambda)$ and $f_2(q,\Lambda)=f_3(q,\Lambda)$, \cite{supp}. Expanding $f_1(q,\Lambda)$ and $f_2(q,\Lambda)$ near $p_1$ \cite{supp} yields $f_1(q,\Lambda)\sim \beta(\Lambda-\Lambda_1)$ and $f_2(q,\Lambda)\sim-i\nu (q-q_0)$. Thus
\begin{eqnarray}
h_q\sim-\imath\nu (q-q_0)\sigma_x+\beta(\Lambda-\Lambda_1)\sigma_z
\end{eqnarray}
with constants $\beta>0$ and $\nu>0$ \cite{supp}. Near the band touching point the effective hamiltonian becomes
\begin{eqnarray}
\mathcal{H}_{\rm{eff}}=\omega_0\sigma_0-\imath\nu(\Lambda) (q-q_0)\sigma_x+m\sigma_z
\end{eqnarray}  
 We identify the mass term $m=\beta(\Lambda-\Lambda_1)$ which cancels out at $\Lambda=\Lambda_1$ at the band crossing point. In the presence of the mass term, the spectrum becomes gapped
 \begin{eqnarray}
\omega_{1,2}\sim \pm\sqrt{(\nu q)^2+m^2}\nonumber
\end{eqnarray} 
Fig.~\ref{fig:f7}(b) shows the magnon speed, (spin wave group velocity) $\nu=\frac{w}{q}$ versus $\Lambda$ around the Dirac point. The jump discontinuities around $\Lambda_1$ and $\Lambda_2$ coincide with the dimerization of the chain, and the limits of the topological phase as shown in Fig.~\ref{fig:f7}(a). This result suggests that the Fermi speed of magnons could be interpreted as susceptibility to $\Lambda$. Its discontinuity would indicate a topological phase transition. 
\section{Discussion}
\label{sec:discussion}
The purpose of this work has been to demonstrate the explicit control of energy bands, magnonic frequencies, and edges modes in arrays of magnetic dipoles. These tunable systems are made out of stacked modulated dipolar chains that interact via magnetic dipolar coupling. When the easy axis anisotropy is strong enough, dipoles settle into parallel magnetic configurations, with gapped 1D and 2D magnonic bands, which are highly tunable by $\Lambda$, the amplitude of the periodic modulation along single chains. The tuning of $\Lambda$ in single chains, gives rise to two topological phase transitions with non-zero Zak phase, which host chiral edge states. In 2D lattices made of modulated chains, the spin-wave volume bands take non-zero Chern numbers $1$ and $-1$, respectively. These values are exchanged at $\Lambda=\frac{3}{2}$ due to two band touchings that yield two monopoles with charge +1 each. Due to the monopoles, the Berry phase acquires divergence, which triggers the exchange of the Chern numbers between the bands. This topological phase transition is manifested in the Hall conductivity of lattices which changes sign at $\Lambda=\frac{3}{2}$. 

We study an effective model for the 1D systems that expose the chiral character of the speed of edge magnons $\nu$. The jump discontinuities in the response of $\nu$ to changes in $\Lambda$, coincide with the two topological phase transitions in chains. Hence, it is proposed that discontinuities in the response function of the magnons Fermi speed could serve as an indicator of topological phase transitions.  

The influence of phonons in the band of magnons have been neglected in this study.  At finite temperatures, phonons could contribute to the thermal Hall current due to the phonon Hall effect \cite{zhang2010topological}. However, it has been argued that this contribution is negligible compared to the magnon Hall current because phonon angular momenta are proportional to the gyromagnetic ratio of charged atoms, which is orders of magnitude smaller than that of magnons \cite{li2021magnonic}.
In the presence of magnon-phonon coupling, the two types of quasiparticles could hybridize near band crossing. Because in two-dimensional materials, thermal agitation would favor the out-of-plane vibrations over the in-plane ones, we expect that the dominant contributions would stem from out-of-plane phonon modes. In our system, the equilibrium magnetic configurations lie in the plane of the sample, and therefore the coupling between magnon and phonon modes should be relatively small. Evidence \cite{li2021magnonic, park2019topological} indicates that in cases when magnon phonon hybridization is realized, it slightly affects the magnon bands in regions of avoided band crossing and that unless the magnon phonon coupling becomes significant, it does not influence the regions of the band where Berry curvature is concentrated.  

The current approach for magnon manipulation leaves out the intervention of external fields and relies instead on tuning a single intrinsic geometrical parameter. The calibration of $\Lambda$ tunes the internal anisotropic magnetic fields in the lattices. Such internal fields originate from dipolar interactions between magnets and produce a spin-momentum locking, which constrains the dipole's magnetic moment orientation, as happens in systems that allow the spin-orbit interaction. The control over these internal fields is accomplished by changing the lattice constant of the system along one single direction, $\Lambda$. 
A thrilling consequence of increasing $\Lambda$ is the flattening of the lower magnon band. The effect of the band flattening on thermal conductivity has not been addressed here, but we believe it deserves special attention. With an intrinsic knob being able to tune the flatness of the spin wavebands, the present dipolar system could open up a new avenue to study correlation-driven emergent phenomena on the background of topological magnonics. 

Possible realizations of the systems presented here could be accomplished by means of molecular magnets \cite{bogani2010molecular,syzranov2014spin}, optical lattices \cite{atala2013direct} and nanomagnetic arrays made out of permalloy \cite{gartside2020current}. Especially suitable are state-of-the-art magnonic crystals fabricated out of epitaxially grown YIG films \cite{frey2020reflection}. Recent work has shown that thickness and width modulated magnonic crystals comprising longitudinally magnetized periodically structured YIG-film waveguides can manipulate magnonic gaps with advantages such as a small magnetic damping and high group velocity of the spin waves \cite{mihalceanu2018temperature}.
%\end{linenumbers}
\section*{Acknowledgments}
We thank  Fondecyt under Grant No. 1210083.

%\bibliography{magnons}
\end{document}